\documentclass[12pt]{article}
\usepackage{epsfig}
\begin{document}
\begin{center}
{\bf \Large\bf Exclusive photoproduction of $f_1(1285)$ meson off proton in the JLab kinematics}
\end{center}
 \vspace{1mm}
\begin{center}
N.~I.~Kochelev, $^{a}$\footnote{kochelev@theor.jinr.ru}
M.~Battaglieri, $^{b}$\footnote{battaglieri@ge.infn.it} R.~De~Vita
$^{b}$\footnote{devita@ge.infn.it} \vskip 1ex \it (a) \it
Bogoliubov Laboratory of Theoretical Physics, Joint Institute for
Nuclear Research, Dubna, Moscow region, 141980 Russia \vskip 1ex
\vskip 1ex  \it $^b$ Istituto Nazionale di Fisica Nucleare, 16146 - Genova, Italy\\

\end{center}
\vskip 0.5cm \centerline{\bf Abstract}
We calculated the exclusive $f_1(1285)$ meson photoproduction
cross section at energy of few GeV within the Regge approach.
The calculation shows that the cross section is sizable, being in the range of 100 nb, and  much
 larger than the the expected cross section of the
$\eta(1295)$ meson photoproduction at the same energy.
These two facts make possible to use this reaction to study the poor known properties
of the $f_1(1285)$  meson in the JLab kinematics.
 \vspace{1cm}

\newpage

\section{Introduction}

Investigation  of hadron properties is nowadays a hot topic,
being subject of several studies
within  non-perturbative QCD approaches.
 The existence  of
possible exotic hadron states is  the subject of both theoretical
\cite{jaffe,Achasov:2008me,Narison:2008nj,Mathieu:2008me} and
experimental activity \cite{Klempt:2007cp,Crede:2008vw,Batta:2009}. The
$f_1(1285)$ meson, with quantum numbers
$I^{G}(J^{PC})=0^+(1^{++})$, is usually considered a member of
 the  axial vector meson nonet. However,
 it was  argued that this resonance may have
a rather large mixture of gluons in its wave function~\cite{fritzsch}.
In Ref.~\cite{km}, the
special role of the $f_1(1285)$ trajectory  in
spin-dependent high energy cross sections, based on the deep
relation of the properties of this meson with the $U(1)_A$ gluon axial
anomaly in QCD, was discussed. The alternative approach to treat
$f_1(1285)$ as a dynamically generated resonance through the
interaction of vector and pseudoscalar mesons in $K^*\bar K$
channel was suggested in Ref.~\cite{oset}.
It is worth to notice  that the
$f_1(1285)$ meson has a large branching ratio ($\sim$36\% \cite{PDG})
to $a_0(980)\pi$. Therefore the production of this meson gives
also a unique opportunity to study the properties of
$a_0(980)$ meson, a well known candidate for exotic
four-quark state (see discussion and references in Refs.~\cite{jaffe} and
\cite{Achasov:2008me}).

Photoproduction is a very
powerful tool to investigate meson properties. The experimental
program of the CLAS collaboration at Jefferson Lab includes various
photoproduction reactions with mesonic final states. In the light
of the importance of the $f_1(1285)$ meson, we
report on an estimate of the  cross section  for the  reaction
$\gamma p \to  p f_1(1285)$ at photon  energy of few GeV, within the Regge approach.

\section{Estimate of the $f_1(1285)$ meson exclusive photoproduction cross section}

\begin{figure}[h]
\vspace{5.cm}
\includegraphics{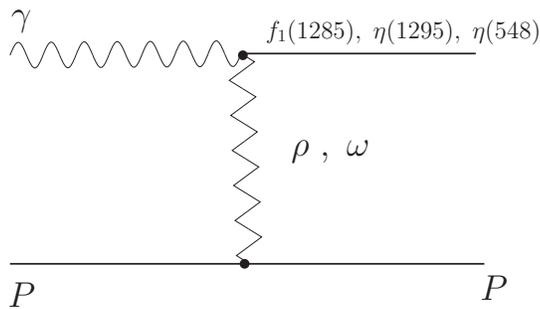}
\caption[]{The dominant diagram in
$f_1(1285)$, $\eta(1295)$, and $\eta(548)$ meson  exclusive
photoproduction off proton within the Regge model.}
\label{fig1}
\end{figure}

The Regge model for meson photoproduction  is being widely used  to calculate   cross sections
for different reactions in the kinematic region $s>>-t$ (see Refs.~\cite{laget},
\cite{sib} and references therein).
Within this approach,  the main contribution to the $f_1(1285)$ photoproduction cross
section  at small  momentum transfer ($-t\leq 1 $
GeV$^2$)  and photon energy range of few GeV    is related to the $t$-channel exchange
of  $\rho$ and $\omega$ meson trajectories (see Fig.~\ref{fig1}).
Propagators of $\rho $ and $\omega $ mesons are given by~\cite{laget}:
\begin{equation}
P_V=(g^{\mu\nu}-\frac{k_V^\mu k_V^\nu}{m_V^2})\big(
\frac{s}{s_0}\big)^{\alpha_V(t)-1}\frac{\pi\alpha_V^\prime}{sin(\pi\alpha_V(t))\Gamma(\alpha_V(t))}D_V(t),
\end{equation}
where $D_V(t)$  is the signature factor. It is well known that
Regge trajectories  can be
either non-degenerate or degenerate~\cite{Collins}.
In Ref.~\cite{Guidal:1997hy}, a  detailed analysis  of high energy pion
photoproduction data within Regge approach was performed. It was
argued that the $\rho$ meson trajectory should be degenerate in order
to describe the ratio of cross sections of charged pions
photoproduction. However, the $\omega$ trajectory should be
non-degenerate to reproduce the dip around $t\approx -0.6$ GeV$^2$
observed in high energy exclusive $\pi^0$ photoproduction.
Using the results of this study, we adopted the following expressions for the
 signature related factors:
\begin{equation}
D_\omega(t)=\frac{-1+exp(-i\pi\alpha_\omega(t))}{2},
\end{equation}
\begin{equation}
D_\rho(t)=exp(-i\pi\alpha_\rho(t)),
\end{equation}
where  $k_V$ is the meson momentum, $s_0=1$ GeV, and ${\alpha^\prime}_V$
 is the slope of the trajectory. For the $\rho$ trajectory the rotating phase
 was chosen~\cite{laget} \footnote{We have checked that the choice of  a constant value for the  phase
  of  the  $\rho$ trajectory
leads to very similar numerical results for the
differential photoproduction cross sections of the $f_1(1285)$, $\eta(1295)$, and $\eta(548)$ mesons.} to be:
\begin{eqnarray}
\alpha_\omega(t)=0.44+0.9t,\\
\alpha_\rho(t)=0.55+0.8t.
\end{eqnarray}
The  vector meson (VM)-proton coupling is given by
the standard expression:
\begin{equation}
{\cal L} =g_V\bar N\gamma_\mu NV_\mu+\frac{g^T_V}{2m_N}\bar
N\sigma_{\mu\nu} NV_{\mu\nu},
\end{equation}
where $V_{\mu\nu}=\partial_\mu V_\nu-\partial_\nu V_\mu $.
The numerical value of the coupling constants were taken from Ref.~\cite{sib}:
\begin{eqnarray}
g_{\omega NN}=10.6,\\
g^T_\omega=0,\\
g_{\rho NN}=3.9,\\
g^T_\rho/g_\rho=6.1.
\end{eqnarray}
The $f_1$-VM-photon vertex  has the following  form \cite{km}:
\begin{equation}
V_{Vf_1\gamma}=g_{Vf_1\gamma}k_V^2\epsilon_{\mu\nu\alpha\beta}\xi^\beta\epsilon_V^\nu\epsilon_\gamma^\alpha
q^\mu, \label{ver}
\end{equation}
where $q$ is photon momentum, $ \xi,  \epsilon_V, $ and $
\epsilon_\gamma $ are the polarization vectors of the $f_1$, the vector meson
and the photon, respectively.

The coupling in Eq.\ref{ver} corresponds to the
$AVV$ Lagrangian obtained in Ref.~\cite{messner}
by using the hidden gauge approach.
We should also mention that this coupling
satisfies the Landau-Yang  theorem \cite{yl} and leads to a
vanishing value  for an  axial vector meson coupling to   two massless
vector particles (e.g. in the limit $k_V^2\rightarrow 0$).

The coupling $g_{\rho f_1\gamma}=0.94$ GeV $^{-2}$  was fixed from the
measured width:
\begin{equation}
\Gamma
_{f_1\rightarrow\rho\gamma}=\frac{m_{\rho}^2(m_{f_1}^2+m_\rho^2)(m_{f_1}^2-m_\rho^2)^3}{96\pi
m_{f_1}^5}g_{\rho f_1\gamma}^2,
\end{equation}
assuming $\Gamma _{f_1\rightarrow\rho\gamma}\simeq 1.3$  MeV \cite{PDG}.

There is no experimental information about  the
$f_1\omega\gamma $ vertex. However, this coupling can be estimated
within the quark model through the known value of $g_{\rho f_1\gamma}$
by using  a quite general flavor decomposition of the $f_1$ wave
function:
\begin{equation}
f_1=\alpha(\bar uu+\bar d d)+\beta\bar s s +\gamma{gg},
\end{equation}
where the parameters $\beta$ and $\gamma$ describe a possible mixture
of strange quark and gluons in $f_1$. In the $SU(2)_f$ limit we
have:
\begin{equation}
g_{\omega f_1\gamma}\approx \frac{e_u+e_d}{e_u-e_d} g_{\rho
f_1\gamma}, \label{ll}
\end{equation}
where  $e_q$ is the electric charge of the correspondent quark.

To further proceed in the   calculation, we need an estimate of  the two  form factors in the
$Vf_1\gamma$ and $VNN$ vertexes.
In the spirit of vector meson dominance,
we  derived $F_{VNN}$ from the Bonn model \cite{sib}:
\begin{equation}
F_{VNN}(t)=\frac{\Lambda_1^2 - m_V^2}{\Lambda_1^2 - t},
\label{form1}
\end{equation}
with $\Lambda_1=1.5$ GeV, and  we chose $F_{Vf_1\gamma}$ in the form:
\begin{equation}
F_{Vf_1\gamma}=\big
(\frac{\Lambda_2^2-m_V^2}{\Lambda_2^2-t}\big)^2 \label{form2},
\end{equation}
with $\Lambda_2=1.04 $ GeV. This form  follows from the recent results
of the $L3$ Collaboration about the  $f_1(1285)$ production in $\gamma\gamma^*$
interaction \cite{L3} and the assumption on  the  similarity of
the heavy photon and vector meson vertexes.
\begin{figure}
\vspace{10.cm} \includegraphics{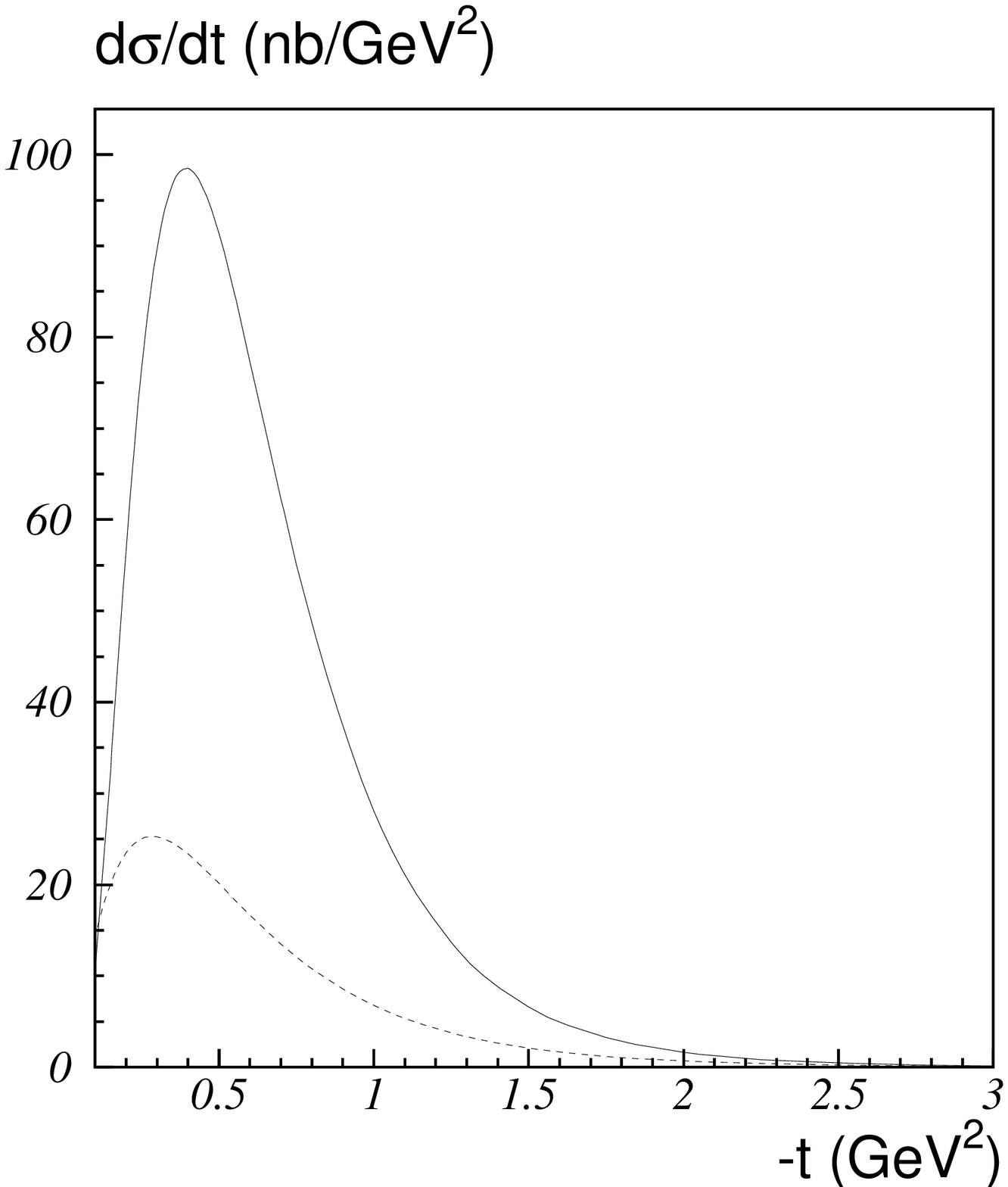} \includegraphics{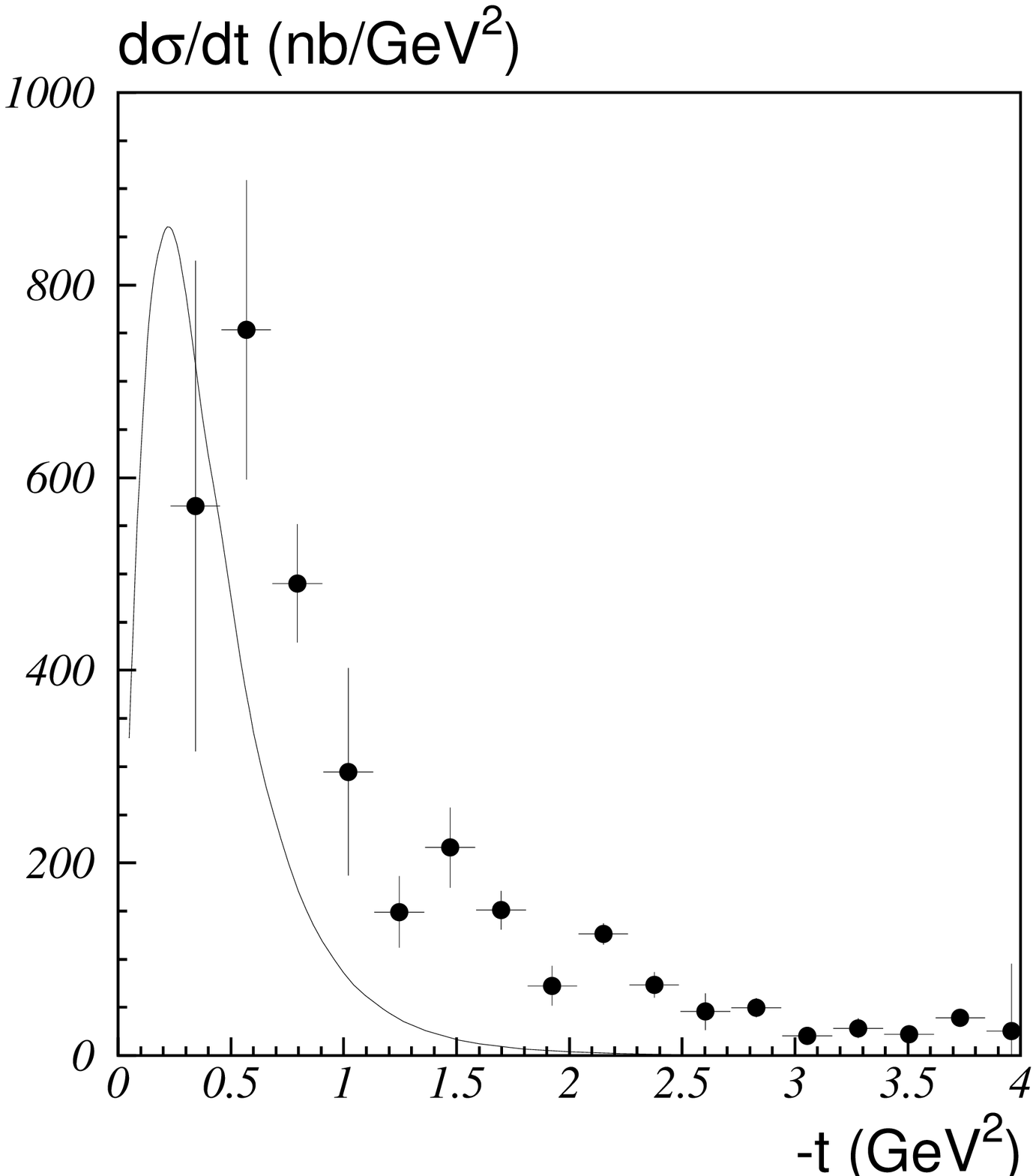}
\caption[]{Estimated cross sections
 for the reactions: $\gamma p \to f_1(1285) p$ (left panel, solid line),
 $\gamma p \to \eta(1295) p$ (left panel, dotted line) and  $\gamma p \to \eta(548) p$ (right panel)
at $E_\gamma=3.1$ GeV.
In the right panel, the SAPHIR data~\cite{SAPHIR} for the  $\eta(548)$ cross section at  $E_\gamma = 2.8-3$~GeV
 are also shown.  }
\label{fig2}
\end{figure}

The resulting differential cross section for
 $E_\gamma=3.1$ GeV is plotted as a solid line  in Fig.~\ref{fig2}-left.
As shown in the plot, the cross section has its maximum
at $-t\sim0.5$ GeV. Both values, $E_\gamma$ and $-t$, are well matched to the kinematics accessible with the
CLAS detector  at JLab.
The size of the cross section, $\sim$100 nb, makes the measurement feasible with such detector.

\section{Estimate of the $\eta(1295)$ exclusive photoproduction cross section}

Experimentally, the main problem to  measure the exclusive
$f_1(1285)$ photoproduction cross section comes from the
background of the $\eta(1295)$ meson. Separation of the two mesons
could be achievable performing a partial wave analysis that
distinguishes the different quantum numbers. On the other hand,
the small production cross section results in low statistics that
limits the accuracy of these analysis. Another approaches is to
extract the cross section through the inclusive measurement of the
reaction $\gamma p \to p X$, where mesons are identified as peaks
in the spectrum of proton missing mass. In the case of the
$f_1(1285)$ and  $\eta(1295)$ meson, their similar
 mass and a  width, makes it practically impossible to distinguish them.
The measurement of the $f_1(1285)$ cross section would be still possible if the  $\eta(1295)$
cross section was found to be  much lower, assuming  therefore, that the observed signal is dominated by the
$f_1(1285)$ meson production.

In Regge theory, the
$\eta(1295)$ meson exclusive photoproduction is described by the same
diagram as for $f_1(1285)$ meson (see Fig.~\ref{fig1}).
In  the  spirit of the vector meson dominance  model,
the $\eta(1295)$ meson photoproduction cross section
can be estimated knowing the  strength of the vertex
$\eta(1295)\rightarrow \gamma \gamma$.
Unfortunately, the are no direct measurements of this width.
We then used an indirect way to estimate the width
relying on the constituent quark model.
Assuming that the $\eta(1475)$ and $\eta(1295)$ mesons
are the first radial excitations of the $\eta^\prime(980)$ and
$\eta(548)$, respectively, we correlated the existing data on
$\eta(1475) \to \gamma \gamma$ width using the  constituent quark model
relationships for two-photon width of pseudoscalar meson~\cite{gerasimov}:

\begin{equation}
\Gamma (0^{-+}\rightarrow 2\gamma)\propto m_{0^{-+}}^3\sum_qe_q^2,
\label{ga2}
\end{equation}
where $\sum_qe_q^2$ represents  the sum of the  electric charges of
quarks in meson, and therefore:

\begin{equation}
\Gamma (\eta(1295)\rightarrow 2\gamma)\approx \frac{\Gamma(
\eta(1475)\rightarrow 2\gamma)\Gamma (\eta\rightarrow
2\gamma)m^3_{\eta^\prime} m^3_{1295}}{\Gamma
(\eta^\prime\rightarrow 2\gamma)m_\eta^3m^3_{1475}}\approx 0.091
KeV \label{ga3},
\end{equation}
where we  used $\Gamma( \eta(1475)\rightarrow 2\gamma )\approx
0.212 KeV$ \cite{PDG} with the assumption  that $K\bar K\pi$ is
the $\eta(1475)$ dominant  decay mode.

 The   $\eta$-VM-photon
vertex   has the following form:
\begin{equation}
V_{V\eta\gamma}=g_{V\eta\gamma}\epsilon_{\mu\nu\alpha\beta}\epsilon_V^\nu\epsilon_\gamma^\alpha
q^\mu k^\nu_{\eta}, \label{ver2}
\end{equation}
where $k_\eta$ is the  $\eta$ meson momentum.
Using Eq.~(\ref{ver2}) and the vector meson dominance model, we obtained the following expression for the
$\rho\eta\gamma$ coupling:
\begin{equation}
g^2_{\rho\eta(1295)\gamma}\approx \frac{96\pi m^3_\rho m^3_\eta
\Gamma( \rho\rightarrow \eta\gamma)\Gamma (\eta(1295)\rightarrow
2\gamma)}{ (m^2_\rho-m^2_\eta)^3{m^3}_{\eta(1295)}\Gamma
(\eta\rightarrow 2\gamma)}\approx 0.0032 GeV^{-2}, \label{ga5}
\end{equation}
where $\eta\equiv \eta(548)$. The $\eta(1295)$-$\omega$ coupling
has been  estimated with a similar equation as in Eq.~\ref{ll}:
\begin{equation}
g_{\omega \eta(1295)\gamma}\approx \frac{e_u+e_d}{e_u-e_d} g_{\rho
\eta(1295)\gamma}. \label{ll2}
\end{equation}

The Brodsky-Lepage form of the transition form factor in the $V\eta\gamma$ vertex was
used~\cite{brodsky}:
\begin{equation}
F_{V\eta\gamma}= \frac{1}{1-t/(8\pi^2f_{PS}^2)}, \label{form3}
\end{equation}
where $f_{PS}$ is the pseudoscalar decay constant, related to the $\Gamma_{\gamma\gamma}$
partial width by:
\begin{equation}
f_{PS}=\frac{\alpha}{\pi}\sqrt{\frac{M_{PS}^3}{64\pi\Gamma_{\gamma\gamma}}}.
\end{equation}

The resulting differential cross section for the reaction $\gamma p \to p \eta(1295)$
at $E_\gamma=3.1$ GeV is plotted as a dotted line  in Fig.~\ref{fig2}-left.
Integrating the two differential cross sections in the whole $-t$ range, we obtained:
\begin{eqnarray*}
\sigma_{f_1(1285)}=68 nb,\\
\sigma_{\eta(1295)}=18 nb.
\end{eqnarray*}

The  $\eta(1295)$   cross section was found to be  smaller (about
25\%) than the $f_1(1285)$ cross section suggesting that the
extraction of the exclusive $f_1(1285)$ photoproduction is
possible without complicated partial wave analysis in the JLab
kinematics.

As a check of the model, we repeated the same calculation to derive the
differential cross section for the exclusive reaction $\gamma p \to \eta(548) p$.
 In this case the  $\eta(548)$-VM-gamma coupling was  obtained using the formula:
\begin{equation}
g_{V\eta\gamma}= \sqrt{\frac{96\pi
M_V^3\Gamma_{V\rightarrow\eta\gamma}}{(m_V^2-{m_{\eta}}^2)^3}}.
\end{equation}

Results of the calculation are shown in Fig.~\ref{fig2}-right compared to the
experimental points for the same reaction measured by the SAPHIR Collaboration
 \cite{SAPHIR} in a similar photon energy range ($E_\gamma = 2.8-3$~GeV).
Data are  described rather well  by our model. The deviation
between theory and experiment, especially at low $-t$, remains
within a factor two and it is typical for such simple
implementation of the Regge theory. More sophisticated models and,
in particular a better treatment of the shape of the form factors,
would result in a better agreement.

\section{Summary}
In summary,  we calculated the cross section for the exclusive
$f_1(1285)$ meson photoproduction off proton above the baryon
resonance region. The chosen kinematics matches  the  typical
Jefferson Lab, Hall-B, photon experiments. Using the Regge model
with some phenomenological input for the unknown parameters, we
obtained a cross section of the order of 100 nb. In the same
framework, we also evaluated the cross section for the exclusive
$\eta(1295)$ meson photoproduction, which represents the main
background for the $f_1(1285)$ meson extraction  from a
photoproduction experiment. The small value we found for such
background  suggests that a measurement of the  $f_1(1285)$
exclusive photoproduction  cross section  with a detector such as
CLAS is possible.

\section{Acknowledgment}
The authors are grateful to S.Gerasimov for useful discussion.
 NK would like to thank  INFN, Sezione di Genova, for the warm hospitality during
this work.

\end{document}